\documentclass[useAMS,usenatbib]{mn2e}
\usepackage{graphicx}
\usepackage{txfonts}

\catcode`\@=11
\def\gsim{\ifmmode{\mathrel{\mathpalette\@versim>}}
    \else{$\mathrel{\mathpalette\@versim>}$}\fi}
\def\lsim{\ifmmode{\mathrel{\mathpalette\@versim<}}
    \else{$\mathrel{\mathpalette\@versim<}$}\fi}
\def\@versim#1#2{\lower 2.9truept \vbox{\baselineskip 0pt \lineskip
    0.5truept \ialign{$\m@th#1\hfil##\hfil$\crcr#2\crcr\sim\crcr}}}
\catcode`\@=12
\def\msun{\hbox{$M_\odot$}}

\def\pn{\par\noindent}
\def\mpb{\medskip\pn$\bullet$\quad}

\begin{document}
\title[Multiple stellar populations in globular clusters]{Origin of
multiple stellar populations in globular clusters and their helium
enrichment}
\author[Alvio Renzini]{Alvio Renzini$^{1}$\thanks{E-mail: 
alvio.renzini@oapd.inaf.it}\\ 
 $^{1}$INAF - Osservatorio
Astronomico di Padova, Vicolo dell'Osservatorio 5, I-35122 Padova,
Italy}

\date{Accepted ... . Received 1....; in original form}
 \pagerange{\pageref{firstpage}--\pageref{lastpage}} \pubyear{2002}

\maketitle
                                                            
\label{firstpage}

\begin{abstract}
The various scenarios proposed for the origin of the multiple,
helium-enriched populations in massive globular clusters are
critically compared to the relevant constraining observations.  Among
accretion of helium-rich material by pre-existing stars, star
formation out of ejecta from massive AGB stars or from fast rotating
massive stars, and pollution by Population III stars, only the AGB
option appears to be viable. Accretion or star formation out of
outflowing disks would result in a spread of helium abundances, thus
failing to produce the distinct, chemically homogeneous sub-populations
such as those in the clusters $\omega$ Cen and NGC 2808. Pollution by
Population III stars would fail to produce sub-populations selectively
enriched in helium, but maintaining the same abundance of heavy
elements.  Still, it is argued that for the AGB option to work two
conditions should be satisfied: i) AGB stars experiencing the hot
bottom burning process (i.e., those more massive than $\sim 3\;\msun$)
should rapidly eject their envelope upon arrival on the AGB, thus
experiencing just a few third dredge-up episodes, and ii) clusters
with multiple, helium enriched populations should be the remnants of
much more massive systems, such as nucleated dwarf galaxies, as
indeed widely assumed.
\end{abstract}

\begin{keywords}
{\it (Galaxy:)} globular clusters: general -- {\it (Galaxy:)} {\bf
globular clusters: individual: $\omega$ Cen, NGC 1851, NGC 2808, NGC
6388, NGC 6441, M54} -- stars: AGB and post-AGB 
\end{keywords}


\maketitle


\section{Introduction}
The recent discovery of discrete, multiple stellar populations within
several among the most massive globular clusters (GC) in the Milky Way
(e.g., Bedin et al. 2004: Piotto et al. 2007) has brought new interest
and excitement on GC research. Further excitement was added by the
realization that some of these stellar populations are selectively
enriched in helium to very high values ($Y\sim 0.38$), without such enrichment
being accompanied by a corresponding increase in the heavy element abundance
of the expected size, if at all (e.g. Norris 2004; Piotto et al. 2005, 2007).

Thus, two main closely interlaced questions arise: how did such
multiple populations form? and, which kind of stars have selectively
produced the fresh helium, without contributing much heavy elements?
Therefore, understanding the origin of multiple populations in GCs
will need to make major steps towards a better knowledge of a
variety of processes such as GC formation, star formation, as well as
of some long standing issues in stellar evolution, e.g., the
asymptotic giant branch (AGB) phase or the effect of rotation in
massive stars.  Moreover, it is quite possible that some of the
massive GCs with multiple populations are the compact remnants of
nucleated dwarf galaxies (Bekki \& Norris 2006), an aspect that widens
even further the interest for this subject. 
All this together makes multiple populations in GCs an
attractive, interdisciplinary subject of astrophysical investigation.

In this paper the main observational evidences are used to constrain
the proposed possible scenarios for the origin of the multiple
populations in GCs, and their associated chemical enrichment
processes. In Section 2 the main relevant observational facts are
briefly reviewed, and in Section 3 AGB stars and fast rotating massive stars
are discussed as possible helium producers, along with
Population III stars and the possible role of deep mixing during the
red giant branch (RGB) phase of low mass stars.  In Section 4 various
proposed scenarios for the origin of the multiple populations are
confronted with the relevant observational facts, arguing that only
massive AGB stars appear to remain viable as helium producers. A
general discussion and the main conclusions are presented in Section
5.

\section{Observational facts}
The main observational facts concerning the evidence for multiple
stellar populations in Galactic GCs are briefly presented in this
section, separately for photometric and spectroscopic observations.
Further collective information and references can be found in a recent
review on these topics by Piotto (2008).

\subsection{Color-Magnitude Diagrams}
\subsubsection{$\omega$ Cen}
It was recognized in the early 'seventies that stars in the GC
$\omega$ Cen span a wide range of metallicities (Cannon \& Stobie
1973; Freeman \& Rodgers 1975) and since then it has been considered
as a unique, exceptional cluster. Over the last several years there
has been a surge of interest on this cluster, starting with the
discovery that its {\it broad} RGB actually resolves into several
distinct RGBs (Lee et al. 1999; Pancino et al. 2000). Then came
the discovery that also its main sequence (MS) splits into two
parallel sequences, with the bluer one being more metal rich by nearly
a factor of two, hence indicating a higher helium abundance (Bedin et
al. 2004; Norris 2004; Piotto et al. 2005).  From the color-magnitude
diagram (CMD) of the subgiant branch (SGB) region we now know that
this cluster includes at least 4, possibly 5 distinct stellar
populations (Sollima et al 2005; Lee et al. 2005; Villanova et
al. 2007). The majority of the cluster stars ($\sim 57\%$) populate
the {\it red} MS, have [Fe/H] = --1.7 and are assumed to have
primordial helium abundance, $Y=0.25$; $\sim 33\%$ belong to the {\it
blue}, helium enriched MS with [Fe/H] = --1.4 and $Y\simeq 0.38$,
which implies a huge helium enrichment ratio $\Delta Y/\Delta Z\simeq
70$ (Piotto et al. 2005).  The residual $\sim 10\%$ of the stars
belong to a metal rich component for which discrepant
spectroscopic estimates exist, ranging from [Fe/H] = --1.1 $\pm 0.2$
(Villanova et al. 2007) to [Fe/H] = --0.6 (Pancino et al. 2002). Part
of the discrepancy may be due to the small number of stars in this
group that have been observed at high resolution.  For these stars 
we have no direct hint on the helium abundance, but values as high as
$Y\sim0.40$ have been proposed (Sollima et al.  2005; Lee et
al. 2005). How these three MS components map into the many SGB, RGB and
horizontal branch (HB) components of this cluster remains partly
conjectural. However, on the RGB there is well established evidence
for sodium and aluminium being anticorrelated with oxygen and
magnesium (Norris \& Da Costa 1995; Smith et al. 2000), indicative
that in a fraction of the stars material is present that was processed
through hydrogen burning at high temperatures. Note that the $\sim
33\%$ overall fraction of the blue MS population takes into account
the observed radial gradient in this fraction (Bellini et al. in
preparation), the helium rich population being more centrally
concentrated.

With a mass of $\sim 3\times 10^6\;\msun$, $\omega$ Cen is the most
massive GC in the Galaxy (Pryor \& Meylan 1993). On this basis, it is
worth estimating the amount of fresh helium and iron that was produced
and is now incorporated in the minority populations. Given its mass
($\sim 10^6\;\msun$) and helium abundance, the intermediate
metallicity, helium rich population includes $\sim 1.4\times
10^5\;\msun$ of fresh helium (i.e., helium of stellar origin) and
$\sim 25\;\msun$ of fresh iron, having adopted $Z^{\rm Fe}_\odot =
0.0013$.  Similarly, the most metal rich sub-population includes $\sim
18\;\msun$ of fresh iron but we cannot presently estimate its helium
enrichment, if any.

\subsubsection{NGC 2808}
Based on the multimodal HB morphology of this cluster, D'Antona \&
Caloi (2004) speculated that NGC 2808 harbours three main populations, each
with a distinct helium abundance. This has been nicely confirmed by
the discovery of a triple MS in this cluster (Piotto et al. 2007), with
$\sim 63\%$ of the stars being assigned the primordial helium
abundance ($Y=0.25$), $\sim 15\%$ having helium enhanced to $Y\sim
0.30$, and $\sim 13\%$ being up to $Y\sim 0.37$. The residual $\sim
10\%$ of the stars are likely to be binaries. Given the narrow RGB
sequence, no iron abundance differences appear to be associated with
these helium differences. However, there appears to be a multimodal
distribution of [O/Fe] ratios among the RGB counterparts of the three
MS populations (Carretta et al. 2006), suggesting that most oxygen has
been turned to nitrogen in the helium-enriched populations.Thus, this
cluster represents an even more extreme case as far as the helium
enrichment parameter $\Delta Y/\Delta Z$ is concerned, as there is no
detectable increase of the overall metallicity in the helium enriched
populations.

The mass of this cluster is $\sim 1.6\times 10^6\;\msun$ (Pryor \&
Meylan 1993), and therefore the intermediate helium-rich, and the
very helium-rich populations contain $\sim 1.4\times 10^4\;\msun$ and
$\sim 2.7 \times 10^4\;\msun$ of fresh helium, respectively, with no
appreciable extra iron.

\subsubsection  {NGC 1851}
No multiple main sequences have yet been detected in the CMD of this cluster,
which clearly shows a double SGB, with the two components
being nearly equally populated ($\sim 55$ and $\sim 45\%$), and
separated by $\sim 0.1$ mag (Milone et al. 2008). If interpreted in
terms of age, this luminosity difference would imply an age
difference of $\sim 1$ Gyr. Alternatively, it has been proposed that
the double SGB may be due to one of the two populations being enhanced
in CNO elements by a factor of $\sim 2$ relative to the other (Cassisi
et al. 2008).  

\subsubsection{M54}

The CMD of this cluster and of the superimposed core of the tidally
disrupted Sagittarius dwarf is extremely complex, with evidence of
multiple MS turnoffs (Siegel et al. 2007). However, there is no
evidence of multiple populations within the cluster itself.

\subsubsection{NGC 6388 and NGC 6441}

Much observational and theoretical efforts have been dedicated to
these two metal-rich bulge clusters, since the discovery that their
HBs exhibit a remarkable blue extension (Rich et al. 1997). Unique
among all GCs, they contain a number of RR Lyraes with an average
period as long as 0.75 days, making them a third Oosterhoff type
(Pritzl et al. 2000). The similarity of their blue HBs with those of
$\omega$ Cen and NGC 2808 has been emphasized by Busso et al. (2007)
who further explore helium enrichment as the cause of their unusual
HB (as originally suggested by Sweigart \& Catelan 1999). Based on CMDs
that include HST ultraviolet bands (F225W and F336W), Busso et al. advocate
a very high helium enhancements (up to $Y=0.40$ in NGC 6388 and $Y=0.35$
in NGC 6441)) for $\sim 15\%$ of the stars in each cluster. A more
moderate helium enrichment ($Y=0.30$ is suggested by Yoon et
al. 2007), but their HB data do not include the very hot extension
revealed by the UV observations.

Both clusters have a mass of $\sim 1.6\times 10^6\;\msun$ (Pryor \&
Meylan 1993), implying that the fresh helium content of the two
clusters is $\sim 3.8\times 10^4\;\msun$ and $\sim 1.4\times
10^4\;\msun$, respectively for NGC 6388 and NGC 6441. Given their low
galactic latitude, these GCs are affected by fairly high reddening
(hence differential reddening) which has so far prevented checking for
the presence of multiple MSs and SGBs as would be expected if the
interpretation of their HB morphology in terms of helium enhancement
is correct. However, this situation may improve soon, and there is
already an indication for a split in the SGB of NGC 6388 (Piotto
2008).

\subsection{Abundance Anomalies}
Departures from chemical homogeneity among stars within individual GCs
are known since a long time (see e.g., the reviews by Kraft 1979, 1999
and Gratton, Sneden \& Carretta 2004). Generally referred to as {\it
abundance anomalies} such departures include a variety of elements and
molecules, such as CN and CH bimodality, the Na-O, Al-O, and Al-Mg
anticorrelations, all indicative of contamination by materials having
been exposed to hydrogen burning at high temperatures ($T>40\times
10^6$ K). Some or all such anomalies are also exhibited by the massive
clusters discussed in the previous section, but to some extent
 affect virtually all well studied GCs, irrespective of their mass.

For example, from the strength of the NH bands among RGB stars in
NGC 6752 Yong et al. (2008) infer than the nitrogen abundance among the
sample stars spans almost two orders of magnitude, with no obvious
bimodality. Moreover, Yong et al. argue that a similar spread must
exist in many other clusters, given that their whole RGB shows a
spread in the NH-sensitive Str\"omgren's $cy$ index of similar size to
that of NGC 6752.  Unfortunately, no similar $cy$ data are presented
for the clusters with multiple stellar populations discussed in the
previous section.  Among them, NGC 2808 exhibits the canonical Na-O
anticorrelation, with three distinct peaks in oxygen abundance
(Carretta et al. 2006), that are likely to be associated with the
three distinct populations revealed by the main sequence photometry
(Piotto et al. 2007). Among the clusters mentioned in section 2.1, NGC
6441 (Gratton et al. 2007) and NGC 6388 (Carretta et al. 2007) exhibit
the Na-O anticorrelation and so does NGC 1851, which exhibits also
Al-O anticorrelation and variations of the s-process elements (Yong \&
Grundahl 2008).

In $\omega$ Cen star-to-star variations of CNO elements exist as well,
but their overall [C+N+O/Fe] abundance ratio appear to be constant
within a factor $\sim 2$, as typical of all GCs (e.g., Pilachowski et
al. 1988; Norris \& Da Costa 1995; Smith et al. 2000, 2005). In particular,
this holds for each of the various metallicity groups in $\omega$ Cen,
which also exhibit large dispersions in the Al abundance (Johnson et
al. 2008)

\subsection{Summary of Observational Constraints}

The main observational constraints that will be used in the following
to narrow down the options on the origin of the multiple stellar
populations in GCs can be summarized as follows:

\mpb All GCs with confirmed (or highly probable) multiple stellar
populations ($\omega$ Cen, NGC 2808, NGC 1851, NGC 6388, and NGC 6441)
belong to the sample of the 10 most massive GCs in the Galaxy (with
$M>10^6\;\msun$).

\mpb Massive GCs with $M>10^6\;\msun$ exist that do not show evidence
for multiple MSs and/or SGBs, nor multimodal HBs (e.g., 47 Tuc,
Sirianni et al. 2005).

\mpb
Some sub-populations can only be understood in terms of high helium content,
up to $Y=0.37$ or more.

\mpb Multiple stellar populations within each GCs are characterized by
discrete values of the helium and iron abundances, i.e., there appears
to be no composition spread within individual sub-populations as the
width of the sequences on the CMDs is consistent with being due only
to known photometric errors. 

\mpb Clusters with helium-enriched multiple populations also tend to
exhibit evidence for Na and Al being anticorrelated with O and Mg,
indicative of materials that have been exposed to hydrogen burning in
a hot environment. However, such variations appear to be virtually
universal among GCs, no matter whether they exhibit multiple main
sequences or not.  Helium enrichment does not appear to be associated
with an increase of [C+N+O/Fe].

\mpb A massive cluster (M54) sits at the center of the core of the
Sagittarius DSph galaxy, and is embedded in its multiple stellar
populations. This offers a concrete example of a massive GC being the nucleus
of a dwarf galaxy.

\section{Helium Producers}
Critical for understanding the origin of the multiple populations in
GCs is the identification of the kind of stars responsible for the
production of the excess helium now incorporated in some of these populations.
Three kinds of stars have been discussed in the literature, namely:
AGB and Super-AGB stars, massive rotating stars, and Population III stars.

\subsection{AGB and Super-AGB Stars}
AGB stars have long been considered for being responsible for at least
 some of the composition anomalies of GC stars (D'Antona, Gratton \&
 Chieffi 1983; Renzini 1983; Iben \& Renzini 1984). Indeed, AGB stars
 present two attractive characteristics, namely: 1) they eject large
 amounts of mass at low velocity ($\sim 10-20$ km s$^{-1}$) which can
 then be retained within the potential well of GCs, and 2) the ejected
 materials can be highly processed through hydrogen burning at high
 temperature, hence being enriched in He and N, and presenting the
 Na-O and Al-O anticorrelations (e.g., Renzini \& Voli 1981; D'Antona
 \& Ventura 2007; Karakas \& Lattanzio 2007). However, three
 difficulties with the AGB scenario have been pointed out: the helium
 abundance, the mass of the secondary populations relative to the
 primary one, and the constancy of [C+N+O/Fe] (e.g., Karakas et
 al. 2006; Karakas \& Lattanzio 2007; Choi \& Yi 2008). These
 difficulties are here addressed again.

Among AGB stars, especially interesting are those in the mass range
$\sim 3-8\; \msun$, because they experience the hot bottom burning
(HBB) process that even in the presence of carbon third dredge-up (3DU)
prevents the formation of carbon stars (Renzini \& Voli 1981), and
will produce the Na-O and Al-O anticorrelations. Instead, in lower
mass AGB stars the C/O ratio largely exceeds unity, and especially so
at low metallicities. If stars were to form from these AGB ejecta they
would be carbon stars, whereas such stars are absent in the clusters
with multiple populations. Moreover, $\sim 3-8\; \msun$ stars
experience the so-called second dredge-up (2DU) shortly before
reaching the AGB (Becker \& Iben 1979), leading to a sizable helium
enrichment in the whole stellar envelope. Besides the 2DU, also the
the 3DU and the HBB may contribute to increase the helium abundance in the
envelope of massive AGB stars, see e.g., Fig. 11 in Renzini \& Voli
(1981). However, Renzini \& Voli models assumed the validity of a
universal core mass-luminosity relation for AGB stars, whereas
Bl\"ocker \& Sch\"onberner (1991) showed that this is no longer valid
once the HBB process operates. Instead, in the presence of HBB the
luminosity of AGB stars increases dramatically, driving stars to very
high (superwind) mass loss rates and leading to an earlier termination
of the AGB phase. Although this scenario is generally accepted, the
precise duration of the AGB phase, hence the extent to which the 3DU
and the HBB process operate in massive AGB stars, all remain highly
model dependent. Thus, a reasonable {\it lower limit} to the amount of
helium enrichment is given by the 2DU contribution alone, which is
fairly well established.

This assumption is consistent with the Bl\"ocker \& Sch\"onberner
(1991) result, which suggests a very prompt ejection of most of the
envelope shortly after the onset of the HBB process.  In practice,
there would be time for the hot CNO processing of these elements
originally present in the star, converting most of C and O into N
(Renzini \& Voli 1981), as well as for establishing the
anticorrelations of Na and Al with O and Mg. These are indeed fairly
rapid nuclear processes once the temperature at the base of the
convective envelope is high enough. Moreover, the prompt ejection of
the envelope drastically reduces the time spent on the thermally
pulsing AGB (e.g., over the early estimates of Renzini \& Voli),
suppressing along with it most of the 3DU events, hence preventing an
appreciable increase of the overall CNO abundance in the envelope.
Thus, it is quite plausible for AGB stars with HBB to eject material
in which helium is highly enriched, CNO nuclei are globally not
significantly enhanced, but have approached their nuclear equilibrium
partition, and anticorrelations among other nuclei have been
established by proton captures at high temperatures. Thus, in
this scenario there is no significant increase of C+N+O due to the
3DU.

In summary, it is assumed here that AGB stars experiencing the HBB
process ($3\lsim M_{\rm i}\lsim 8\;\msun$) 1) eject the whole envelope
shortly after the onset of HBB, 2) the ejecta are enriched in
helium solely by the 2DU, 3) too few 3DU events have time to take
place, and no appreciable increase of the overall CNO abundance
occurs in the envelope, and 4) HBB is sufficiently effective to
promptly establish the Na-O and Al-O anticorrelations. Some
observational evidences support these assumptions.  The mass
distribution of white dwarfs terminates at $\sim 1.1\;\msun$
(Bergeron, Saffer \& Liebert 1992; Bragaglia, Renzini \& Bergeron
1995; Koester et al. 2001), indicating that the core mass of AGB stars
does not grow beyond this limit. This also implies that there must be
very little, if any, increase of the core mass during the evolution of
the most massive AGB stars, since the core mass of a $\sim 8\;\msun$
star just after completion of the 2DU is already  $\sim
1.1\;\msun$ (Becker \& Iben 1979). In addition, an AGB calibration
based on globular clusters in the Magellanic Clouds indicates that in
clusters younger than $\sim 300$ Myr there is negligible contribution
of AGB stars to the bolometric luminosity of the clusters (Maraston 2005).
This implies a very short AGB phase and a very small, if any, increase of
the core mass during the AGB phase of stars more massive than $\sim
3\;\msun$. 

Assuming that the first stellar generation formed in a short,
virtually instantaneous burst, $\sim 3$ to $8\;\msun$ stars were
shedding their envelope between $\sim 30$ and $\sim 300$ Myr after the
burst. It is during this time interval that helium-enriched AGB ejecta
may have accumulated inside the cluster potential well, and new stars
may have formed out of them. The amount of fresh helium released by
stars of initial mass $M_{\rm i}$ assuming only the 2DU contribution
can be easily estimated from Becker \& Iben (1979), see also Fig. 1 in
Renzini \& Voli (1981), where the mass of dredged-up helium is $\sim
0$ for $M_{\rm i}=3\;\msun$ and increases linearly to $\sim 1\;\msun$
for $M_{\rm i}=8\;\msun$. Note that only 3/4 of the dredged-up
helium is ``fresh'', i.e., has been synthesized within the star
itself, whereas 1/4 is primordial, given that $Y=0.25$ in the first
stellar generation. Therefore:
\begin{equation}
\Delta M_{\rm He}\simeq 0.15(M_{\rm i}-3)\;\; \msun,
\end{equation}
which applies to a metal-poor population with $Z=0.001$. Here only
stars with initial mass in the interval $3<M_{\rm i}<8\;\msun$
experience the 2DU before the AGB phase, then soon activate the HBB
process and expel the envelope according to the scenario sketched
above.  Thus, convolving the mass of fresh helium with the initial
mass function (IMF) one can derive the total mass of fresh helium
produced and expelled by these 3 to 8 $\msun$ stars per unit stellar
mass in the whole population. For a Salpeter IMF (slope $x=1.35$) for
$M_{\rm i}>0.5\;\msun$, and a flatter IMF for the lower mass stars
with $x=0.35$ for $M_{\rm i}<0.5\;\msun$, one then derives:
\begin{equation}
 M_{\rm He} \simeq 0.007\times M_{\rm tot},
\end{equation}
 i.e., the mass of fresh
helium released is $\sim 0.7\%$ of the original mass of the parent 
stellar population.

In a similar fashion, one can estimate the helium abundance $Y$ in the
ejecta of the 3 to 8 $\msun$ stars, which ranges from the primordial
value $Y=0.25$ for $M_{\rm i}=3\;\msun$ to $Y\simeq 0.36$ for $M_{\rm
i}=8\;\msun$. Integrating over the same IMF, one gets that the average
helium abundance of the ejecta from 3 to 8 $\msun$ stars is
$<\!Y\!>=0.31$. Higher values could be obtained by restricting the
integration over a narrower mass range, e.g., 4 to 8 $\msun$ or 5 to 8
$\msun$.  For the minimum mass approaching 8 $\msun$ then $<\!Y\!>$
tends to 0.35, but the total amount of released fresh helium
would vanish.

Stars in the range $8\lsim M_{\rm i}\lsim 10\;\msun$ ignite carbon
non-explosively and may leave O-Ne white dwarfs, or proceed to
electron-capture/core-collapse supernovae. The former outcome results
if the envelope is lost in a (super)wind during helium-shell burning
(the Super-AGB phase). Conversely, a supernova explosion ends the life of
the star if mass loss during the Super-AGB phase is insufficient to
prevent the core from growing in mass until it finally collapses
(e.g., Nomoto 1984; Ritossa, Garcia-Berro \& Iben 1996,1999;
Poelarends et al. 2008). S-AGB stars have also experienced the 2DU and 
their envelope has been enriched in helium to $Y\simeq 0.38$, which makes them
attractive helium contributors in the context of helium-rich populations in 
globular clusters (Pumo, D'Antona \& Ventura 2008). Assuming that all
8 to 10 $\msun$  S-AGB stars leave O-Ne white dwarfs, one can estimate that 
the fresh helium mass produced by 3 to 10 $\msun$ stars would be $\sim 20\%$ 
higher than given by Eq. (2), i.e.,
\begin{equation}
 M_{\rm He} \simeq 0.009\times M_{\rm tot}.
\end{equation}
The accuracy of this theoretical estimate is difficult to assess. The
helium mass could actually be somewhat higher if the duration of the
AGB phase is long enough to allow the 3DU and HBB processes to
increase the envelope helium beyond the value reached after the 2DU,
but it could be lower if the mass range of the useful S-AGB stars is
narrower than adopted here. (Note that in  both cases the main uncertainty
comes from what one is willing to assume for the mass loss during the
AGB/S-AGB phases.)  Still, one can regard this estimate as quite
reasonable, given our current understanding of AGB/S-AGB evolution.

As far as the average helium abundance is concerned, from the
inclusion of S-AGB stars one predicts a modest increase from
$<\!Y\!>=0.31$ to $<\!Y\!>\simeq 0.33$, or a little higher if
considering a minimum contributing mass somewhat in excess of
$3\;\msun$. In any event, $<\!Y\!>=0.38$ can be regarded as the
upper limit given the assumed AGB/S-AGB evolution.

Finally, it is worth noting that stars below $\sim 3\;\msun$ do not
experience the 2DU and the HBB processes, spend a long time on the AGB
in its thermally-pulsing phase, and experience repeated 3DU episodes
which increase both helium and carbon in their envelopes. The fact
that helium is accompanied by carbon enhancement makes these lower
mass AGB stars less attractive helium producers, because the helium
rich populations in GCs do not appear to be
enriched in carbon (Piotto et al. 2005). Thus, the interesting mass
range is $\sim 3$ to $\sim 10\;\msun$.

In summary, it is assumed that metal poor ($Z\sim 0.001$) AGB stars
more massive than $\sim 3\;\msun$ and S-AGB stars experience the 2DU
and the HBB process, and leave enough for the HBB process to convert
most of the original carbon and oxygen into nitrogen, but not enough
to experience a sufficient number of 3DU episodes to significantly
increase the overall CNO abundance in the envelope. In this respect,
the schematic AGB evolution adopted here differs from AGB
models existing in the literature (e.g., Renzini \& Voli 1981; Groenewegen \&
de Jong 1993; Herwig 2004; Izzard et al. 2004; Ventura \& D'Antona
2005, 2008; Marigo \& Girardi 2007).

\subsection{Massive Rotating stars}

Massive stars also produce sizable amounts of fresh helium, especially
during their Wolf-Rayet phase. However, they also produce metals in
large quantity, whereas the helium rich populations in GCs are very
modestly enriched in metals ($\omega$ Cen), or not at all (NGC 2808).
In the attempt to overcome this difficulty, Maeder \& Meynet (2006)
have proposed fast-rotating massive stars as potential helium
producers in young GCs, a scenario further developed by Decressin
et al. (2007), Decressin, Charbonnel \& Meynet (2007) and Meynet,
Decressin \& Charbonnel (2008). Massive rotating stars would harbor
meridional circulations bringing to the surface products of hydrogen
burning (i.e., helium), while losing (helium enriched) mass in three
distinct and physically separated modes: 1) a slow outflowing
equatorial disk whose helium abundance increases as evolution
proceeds, 2) a regular, fast, radiatively driven wind also enriched
in helium in the directions unimpeded by the disk, and 3) a final
core-collapse supernova explosion. Only the slow outflowing disk is
considered of interest for the production of helium-enriched stars,
because both radiative winds and supernova ejecta run at thousands of
km/s, and would not be retained inside the relatively shallow
potential well of the proto-cluster.

While physically plausible, this scenario suffers from the difficulty
of predicting with any degree of confidence the efficiency of
meridional circulations to mix helium into the stellar envelope, and
the rate of mass loss via the outflowing disk, none of which can be
estimated from first principles. Moreover, star formation would have
to be confined within the outflowing disks around individual stars,
before such disks are destroyed by the fast winds and supernova
ejecta from nearby cluster stars, and mixed with them (Meynet et al. 2008).

\subsection{RGB Self-Enrichment}
In an early attempt to account for some GC abundance anomalies,
Sweigart \& Mengel (1979) proposed that in low mass upper RGB stars
($M\lsim 2\;\msun$) mixing could extend below the formal boundary of
the convective envelope, and reach well into the hydrogen burning shell. Thus,
materials processed by hydrogen-burning reactions within the shell
could be brought to the surface, changing the C:N:O proportions, and
possibly establishing some of the abundance (anti)correlations typical
of GC stars. Moreover, along with processed CNO elements, some helium
enrichment could also take place, hence affecting the subsequent HB
evolution (Sweigart 1997; Sweigart \& Catelan 1998; Moehler \&
Sweigart 2006).

Observations of upper RGB stars in several GCs suggests indeed that
the Sweigart-Mengel process may be at work in these stars, given that
the $^{12}$C/$^{13}$C ratio has reached close to its nuclear
equilibrium value ($\sim 3.5$) in virtually all of them (Recio-Blanco
\& de Laverny 2007). However, several abundance anomalies extend to
much lower luminosities and down to the main sequence (e.g. Gratton et
al. 2004), which means that RGB self-enrichment cannot be the only
process at work. In particular, it cannot account for the
helium-enriched MS stars, unless one is willing to consider the
possibility of forming the subsequent stellar generation out of the
ejecta from low-mass RGB stars. There are several difficulties
with this option, such as the small mass lost by stars that may
experience the Sweigart-Mengel process (i.e., those in the range $\sim
1$ to $\sim 2\,\msun$), relative to AGB/S-AGB stars, or the very long
(several Gyr) time required to accumulate any sizable amount mass
before being suddenly turned into stars. This very long accumulation
time would imply the secondary populations to be several Gyr younger
than the first generation. According to Villanova et al. (2007) there
is actually a hint for the helium rich population in $\omega$ Cen
being a few Gyr younger, but other interpretations of the data exist
that do not require such large age differences (Sollima et al. 2005;
Lee et al. 2005). Moreover, a several Gyr long accumulation time looks
quite unlikely, given the possible interaction of the cluster with the
galactic environment (e.g. disk crossing, tidal stripping, etc.).
Perhaps more fundamentally, metal poor $\sim 1-2\,\msun$ stars become
carbon stars on the AGB, the phase during which most of their mass
loss takes place, and therefore they are not suited to provide raw
material with the proper chemical composition for the production of
secondary populations.  For all these reasons RGB stars are considered
less likely helium producer candidates. Still, RGB self enrichment may
add further variance on top of other processes.

\subsection{Population III Stars}
Finally, in order to achieve very high $\Delta Y/\Delta Z$ values
stochastic contamination by helium produced by Population III stars
was suggested by Choi \& Yi (2007) in a scenario in which Population
III and Population II star formation temporally overlap. Here helium
would be produced by very massive Pop. III stars, and metals by
massive Pop. II stars, and mixing their products in various
proportions one could achieve high values of $\Delta Y/\Delta
Z$. Although this process may take place in nature, it does not appear
to work for explaining the properties of multiple stellar populations
in globular clusters. For example, in the case of NGC 2808 we have
three distinct values of the helium abundance, with the same
metallicity, a pattern that cannot be reproduced by mixing gas that
formed the dominant cluster population with helium-rich gas from
Pop. III stars. Enrichment in helium would indeed be accompanied
by dilution of metals, and the helium rich population would be metal
poor, contrary to the case of both $\omega$ Cen and NGC 2808.

A variant of this scenario was proposed by Chuzhoy (2006), with
Population III stars being pre-enriched in helium thanks to
gravitational settlement of this element within dark matter halos in
the early universe. While this mechanism may produce even higher
helium abundances in the ejecta of massive Population III stars, it is
prone to the same difficulties of the Choi \& Yi scenario if intended
to account for the helium rich populations in GCs.

\section{Scenarios for the origin of the multiple populations}

A most stringent constraint on scenarios for the origin of multiple
populations is their very nature, i.e., being indeed, multiple,
discrete populations each characterized by a specific helium
abundance, as most clearly evident in the case of $\omega$ Cen and NGC
2808. This immediately implies that helium enrichment of the
interstellar medium (ISM) and star formation for the second (and
third) stellar generation were not concomitant processes, but 
instead took place sequentially. Helium-rich material was accumulated
in the ISM (and mixed therein) for a sufficiently long time until
suddenly a burst turned a major fraction of the ISM into
stars. In fact, a continuous star formation process proceeding along
with the ISM helium enrichment would have resulted in a continuous
distribution of helium abundances in the newly formed stars, hence in
a broadening of the main sequence rather than in well separated
sequences as actually observed.  In this Section one tests various
proposed scenarios against this important constraint.

\subsection{Accretion on Pre-existing Stars}

Any gas matter lost by the first stellar generation (e.g., by its AGB
stars) was necessarily less massive than the parent cluster, and hence had
a lower mass density compared to the stellar component if spatially
distributed in roughly the same way. A simple calculation shows that
under such circumstances any stellar mass size portion of the ISM
already included several lower main sequence stars. At first sight
accretion onto these pre-existing seeds would seem more likely than
starting new stars from scratch. Hence, already a long time ago it was
favored as a likely mean to produce chemical anomalies in GCs (e.g.,
D'Antona, Gratton \& Chieffi 1983; Renzini 1983; Iben \& Renzini
1984), and has been considered even recently in the context of the
helium rich populations (Tsujimoto, Shigeyama \& Suda 2007; Newsham \&
Terndrup 2007). However, accretion depends on mass, velocity and orbit
of each star inside the cluster, which would have resulted in a broad
range of accreted masses from the ISM whose helium abundance was
secularly evolving. After Rayleigh-Taylor mixing of accreted matter
with the underlying layers of lower molecular weight, stars would now
show a range of helium abundances. Thus, accretion would inevitably
result in a broad, continuous distribution of stellar helium
abundances, contrary to the required multiple discrete values.  In
spite of its attractiveness, accretion must therefore be rejected as
the primary process having produced the helium-enriched populations, at least
in the two clusters with clearly defined multiple main sequences.

\subsection{Multiple Star Formation Episodes} 
This leaves multiple star formation episodes as the only mechanism
able to produce successive stellar generations, each with a uniform
helium (and metal) abundance. This means that helium-enriched gas had to
accumulate in the potential well of the cluster, without experiencing
any star formation, until something suddenly triggered the burst of
star formation. Yet, it is unlikely that the efficiency of converting
ISM gas into stars was near unity, and the lower this efficiency, the
more massive had to be the first stellar generation in order to
account for the mass of the second generation and its helium content.
Any gas not converted into stars was soon lost from the proto-cluster,
blown in a wind powered by high velocity stellar winds from massive
stars and supernova explosions.

\subsubsection{From AGB/S-AGB Ejecta}
It is reasonable to assume that the first cluster stellar generation
arose from a burst of star formation, of duration shorter than the
lifetime of the most massive stars (i.e., $\lsim 1$ Myr). Then the
subsequent 20--30 Myr were dominated by massive stars, whose winds and
supernovae cleared the protocluster of any residual gas. After the
last core-collapse supernova, i.e., 20--30 Myr since the beginning,
conditions were finally established favoring the low-velocity winds
from S-AGB/AGB stars to accumulate. At the beginning of the
accumulation the helium abundance of such material was that pertaining
to the ejected envelopes of $\sim 8-10\;\msun$ stars, or $Y\simeq
0.38$ according to the estimate presented in Section 3.1. Later, as
AGB stars of lower and lower mass were ejecting their envelope,
contributing material less enriched in helium, the helium abundance in
the ISM kept steadily decreasing, down to $Y\simeq 0.31$ some 300 Myr
after start. Subsequently, as the AGB started to be populated by
$\lsim\; 3\;\msun$ stars, addition of fresh helium from 3DU replaced
that from 2DU, but along with it also carbon and s-process elements
began to accumulate.

Clearly, the time interval between 20--30 Myr and $<300$ Myr is the
most propitious epoch for producing subsequent stellar generations
with an helium abundance and overall composition close to that
demanded by the observations.  More demanding is, however, the
requirement of producing a second or a third generation of the
observed mass. For example, in the case of $\omega$ Cen, according to
Eq. (3) its present first generation of $\sim 2\times 10^6\;\msun$
produced only $\sim 1.6\times 10^4\;\msun$ of fresh helium, even
allowing for a full S-AGB contribution, a factor of $\sim 10$ less than
the observational requirement estimated in Section 2.1.1. We encounter
here the main difficulty of the AGB scenario, as already pointed out
by several authors (e.g., Karakas et al. 2006; Bekki \& Norris 2006;
Karakas \& Lattanzio 2007), i.e., requiring a first generation of
stars much more massive than that still harbored by the cluster.  This
difficulty is further exacerbated if the actual efficiency of star
formation is less than unity.

However, the helium abundance in AGB/S-AGB stars $0.31\lsim Y\lsim
0.38$ is not in sharp contrast with the observational requirements
from the two cases with well established multiple main sequences
($\omega$ Cen and NGC 2808), where in both cases $Y\le 0.38$. Thus,
AGB/S-AGB stars remain viable candidates, though we need much more of
them than the AGB/S-AGB share of the first generation we now see in
these clusters.

One may think that one way of having more AGB/S-AGB stars is by making
recourse to a flat IMF. However, this may work only if the IMF of the
subsequent generations is different from that of the first one, and in
particular much steeper than it: a very contrived scenario. Otherwise,
if the IMFs are the same, with a flat IMF one gets more helium from
the first generation, but by as much increases the amount of fresh
helium demanded by the second generation, and the discrepancy by a
factor of $\sim 10$ remains.  Thus, a flat IMF does not solve the
problem.

\subsubsection{From Massive Rotating Stars}
It is very difficult to prove or disprove those models of fast
rotating massive stars (FRMS) in which helium is mixed in the stellar
envelopes by meridional circulations, and slowly lost in an outflowing
disk. Thus, I take the FRMS scenario as described in Decressin et
al. (2007) at face value. One problem with this scenario is that it
assumes that such disks survive long enough to produce new stars, in
spite of their impervious environment in which they are bombarded from
all directions by fast stellar winds and supernova explosions.  But
even admitting that disks manage to deliver new stars, such stars will
reflect the helium abundance of the disks they are born from, which
varies from one massive star to another, and for any given stars
varies as a function of its evolutionary stage. Thus, stars born out
of such disks will inevitably show a spread in helium abundances, and
one would have broad GC main sequences rather than multiple ones as
demanded by the observations. The existence of multiple, discrete
stellar populations in $\omega$ Cen and NGC 2808 rules out this FRMS
scenario as a viable one to explain the helium rich populations.

Nevertheless, it is possible that meridional circulations are at work
in massive stars, and bring fresh helium to the surface. If so, some
meridional mixing and helium enhancement may not be confined to the
very massive stars exploding as supernovae. Some helium enrichment
might also take place during the main sequence phase of stars
less massive than $\sim 8-10\;\msun$, which will later deliver such
helium during their AGB/S-AGB phase. Therefore, if meridional mixing
of helium exists in $\lsim 10\;\msun$ main sequence stars, it would
alleviate the difficulty for the AGB/S-AGB scenario outlined in the
previous subsection, in particular concerning the helium
abundance, yet by an amount that is hard to guess
theoretically. Spectroscopic observations of $\lsim 10\;\msun$
stars appear to be the only way to assess whether
helium enrichment does indeed take place, either directly from 
helium line strengths in hot stars, or indirectly from C:N:O ratios
indicative of deep mixing.

\subsection{Metal enrichment in $\omega$ Cen}

Contrary to the case of NGC 2808, in $\omega$ Cen the
helium-enriched population identified by the blue main sequence is
also enriched in iron, and contains some $25\;\msun$ of fresh iron
that was not initially present in the first stellar population
(cf. Section 2.1.1). If coming from relatively prompt Type Ia
supernova events, each contributing $\sim 0.7\;\msun$ of iron (e.g.,
Iwamoto et al. 1999), at least $\sim 35$ Type Ia supernovae (SNIa)
from the first stellar generation had to explode within the helium
enriched ISM, and do so within the first $\sim 10^8$ yrs.  This looks
quite plausible given the wide variety of distributions of SNIa delay
times that theoretical models can generate (e.g. Greggio
2005). However, if the excess iron were produced by SNIa's, then the
secondary population would have lower $\alpha$-element to iron ratios,
being selectively enriched only in iron. This is at variance with the
observed [Ca/Fe], [Mg/Fe] and [Si/Fe] ratios in $\omega$ Cen RGB
stars, which show no dependence on the iron abundance (Norris \& da
Costa 1995; Smith et al. 2000; Pancino et al. 2002). Thus, this
excludes SNIa's for being responsible for the iron enrichment, and
favours core collapse supernovae (CCSN), which along with iron produce
also $\alpha$ elements. On average, each CCSN produces $\sim
0.07\,\msun$ of iron (Hamuy 2003), roughly 10 times less than each
SNIa. Thus, a few hundred CCSNe are needed to produce the 25 $\msun$
of iron in the helium-rich population of $\omega$ Cen.

With the adopted IMF, one CCSN is produced every $\sim 100\,\msun$
of gas turned into stars, and therefore with its $\sim 2\times 10^6\,\msun$
the first stellar generation in $\omega$ Cen has produced $\sim
20,000$ CCSNe. If the progenitor was at least 10 times more
massive than the present cluster (see below), then over $2\times 10^5$
CCSNe had been produced. Thus, it is sufficient that $\sim 0.1\%$
of their ejecta were trapped inside the protocluster while mass lost
by S-AGB/AGB stars  had already started to accumulate. A very small
time overlap between CCSN events and the appearance of S-AGB/AGB stars is
necessary for this to happen, consistent with the short timescale
($\lsim 1$ Myr) postulated for the formation of the first stellar
generation (cf. Section 4.2.1).

Note that these supernova explosions should have avoided to trigger
any major star formation, otherwise a continuous distribution of
helium and iron abundances would have resulted.  What triggered the
major star formation burst leading to the second generation remains
unidentified. No attempt is made here to speculate on the full star
formation history in $\omega$ Cen, which is much more complex given
the identification of its 5 sub-populations (Sollima et al. 2005; Lee
et al. 2005; Villanova et al. 2007).

\section{Discussion and Conclusions}

In the previous sections it has been argued that only the AGB/S-AGB
scenario remains viable to account for the helium-enriched
sub-populations.  Still, with a serious difficulty to overcome, plus
some minor ones. If our current understanding of the helium enrichment
in intermediate mass stars is not grossly incorrect, then the mass of
the first stellar population in a cluster such as $\omega$ Cen had to
be several times larger than the present mass of the
cluster. Following Bekki \& Norris (2006) this difficulty can be
solved if clusters such as $\omega$ Cen and NGC 2808 are the remnant
nucleus of a nucleated dwarf galaxy that was torn apart by the tidal
field of the Galaxy. Hence the parent first population providing the
necessary raw material for the successive generation(s) would have
been much more massive than these clusters are today. Some
circumstantial evidence for this now widely entertained scenario is
the finding that M54, one of the most massive GCs in the Galaxy, is
indeed associated with the Sagittarius dwarf, albeit there is no
evidence for multiple populations {\it within} this cluster
(Siegel et al. 2007). The nucleated dwarf NGC
205 may be another example relevant to this scenario: with its tidal
stream towards M31 (McConnachie et al. 2004) it may represent an early
stage of the process suggested by Bekki \& Norris. Its nucleus is
dominated by old stellar populations, and yet it is quite bright in
the WFPC2 ultraviolet F225W and F185W passbands (Cappellari et
al. 1999). If the UV light comes from an extended HB (such as that of
$\omega$ Cen and NGC 2808) it may also harbor helium-enriched
populations, and would represent a possible testbed for the nucleated
dwarf scenario of Bekki \& Norris (2006).

Of course, for the nucleated dwarf scenario to work, successive stellar
populations must have a much lower probability of being tidally
stripped compared to the first population, otherwise the mass
discrepancy would remain. This can only be achieved if the successive
starbursts are far more centrally concentrated compared to the first
one, i.e., the AGB/S-AGB ejecta from the first generation should 
collapse to the very bottom of the potential well before leading to
star formation. In this connection, it is quite reassuring that the
helium-enriched main sequence stars in $\omega$ Cen are indeed
markedly more centrally concentrated than the others (cf. Section
2.1.1).

Besides reproducing the required helium enhancement, a successful
theory for the origin of multiple stellar populations in massive GCs
should also account for the observed abundance and distribution of
CNO and other intermediate mass elements. Several authors (e.g.,
Karakas et al. 2006; Romano et al. 2007; Choi \& Yi 2008) have shown
that yields of existing AGB models fail to satisfy this constraint,
as along with helium stellar ejecta would be enriched also in carbon
from the 3DU. The question is therefore as to whether one should
conclude that the fresh helium does not originate from AGB stars, or
that the used theoretical yields are not correct. Given the current
uncertainties affecting the massive AGB models (especially due to the
treatment of mass loss and convective overshooting) it seems worth
keeping a pragmatic approach, hence taking existing AGB models with
special caution. The schematic AGB evolution described in Section 3.1
is quite plausible given our ignorance of mass loss in bright AGB/S-AGB
stars with HBB, and does not conflict with existing observations.
Actually, it allows keeping AGB/S-AGB stars as viable helium producers for the
multiple stellar generations in globular clusters.

For the present scenario to work it is essential that the AGB ejecta
accumulate for a fairly long time ($\sim 10^8$ yr) without any 
significant star formation before suddenly a major fraction of the
interstellar medium is turned into stars by a burst. This may not be
such an {\it ad hoc} assumption, given that star formation in sporadic
bursts appears to be the norm for dwarf galaxies (Gerola, Sneden \&
Sulman 1980), as also demonstrated by the discrete multiple
generations in dwarfs such as Carina (e.g., Monelli et al. 2003).

In the case of NGC 2808 there had to be a second and a third stellar
generation. If the scenario presented in this paper is basically
correct, then the most helium rich secondary sub-population would have
been the first to form out of the most massive AGB/S-AGB ejecta, which
are the most helium rich. Then the ISM was replenished again by the
ejecta of the less massive AGB stars from the first generation, plus
perhaps a contribution by the secondary generation. If so, then the
generation identified by the {\it middle} of the three main sequences
in this cluster was the last to form.

Worth briefly mentioning are those observational studies that may help
testing the plausibility of the AGB/S-AGB scenario proposed in this
paper. Some of these tests concern the adopted AGB/S-AGB evolution and
nuclear yields, in particular concerning $\sim 3$ to $\sim 10\;\msun$
stars. It would be interesting to test whether the surface helium and
nitrogen abundance are enhanced in $\sim 6$ to $\sim 10\;\msun$ main
sequence stars, possibly by the meridional circulation process
advocated by Maeder \& Meynet (2006), albeit high rotational
velocities may hamper accurate abundance determinations, and massive
stars a metal poor as stars in $\omega$ Cen and NGC 2808 are not
within reach. Direct observations of bolometrically very luminous
AGB/S-AGB stars in the Magellanic Clouds with very high mass loss
rates should help understanding the crucial, final evolutionary stages
of intermediate mass stars.  Moreover, high resolution spectroscopy of
very large samples (several hundreds) of stars in the various
sub-populations in globular clusters should help identify unequivocal
chemical signatures of the {\it donors} of the materials out of which
these sub-populations have formed. These kind of studies are now
possible thanks to the high multiplex multiobject spectrographs at
8--10m class telescopes, such as e.g., FLAMES at the VLT (Sollima et
al. 2005; Carretta et al. 2006; Villanova et al. 2007).  Finally, the
multifrequency study of nucleated dwarfs in and around the Local Group
may help testing whether the most massive globular clusters may have
originated from the tidal stripping of these objects.

In conclusion, excluding scenarios that {\it qualitatively}
conflict with observations, such as accretion, fast rotating massive
stars or Population III stars, turn out to be much easier than proving
others that qualitatively appear to work, but may have {\it
quantitative} difficulties. This is the case for the AGB/S-AGB
scenario advocated in this paper, where the predicted helium abundance
in the secondary populations admittedly falls a little short of the highest
values suggested in the literature ($Y=0.38-0.40$). In this respect,
one should consider than helium abundance estimates are affected by
uncertainties that must be of the order of a few 0.01, hence no
macroscopic discrepancy appear to exist. Still, it would help if,
besides the 2DU, other processes contribute a little additional fresh
helium in $\sim 3$ to $\sim 10\,\msun$ stars. Additional helium may
come from the HBB process operating during the AGB/S-AGB phase, and/or
from meridional circulations during the main sequence phase of the
progenitors of AGB/S-AGB stars.

\section*{Acknowledgments}
I would like to thank Giampaolo Piotto for numerous stimulating
discussions on the multiple populations in globular clusters and for a
critical reading of the manuscript. Useful suggestions on the metal
enrichment by supernovae are acknowledged from Laura Greggio. Finally,
I would like to thank the anonymous referee for his/her many constructive
comments.


\label{lastpage}

\end{document}